\documentclass{acm_proc_article-sp}
\usepackage{graphicx}
\usepackage{textcomp}
\usepackage{fancyvrb}
\usepackage{listings}
\usepackage[unicode=true]{hyperref}
\usepackage{enumitem}

\newcommand{\sdlrx}{SDL-R$_X$}

\begin{document}

\conferenceinfo{}{Bloomberg Data for Good Exchange 2016, NY, USA}

\title{Internet Scale Research Studies using \sdlrx{}}

\numberofauthors{2}
\author{
\alignauthor
James Kizer, Arnaud Sahuguet\\
       \affaddr{The Foundry @ Cornell Tech}\\
       \affaddr{Cornell Tech, New York USA}\\
       \email{\{jdk288, arnaud.sahuguet\}@cornell.edu}
\and
Neil Lakin, Michael Carroll, JP Pollak, Deborah Estrin
\\
       \affaddr{Small Data Lab @ Cornell Tech}\\
       \affaddr{New York City, NY}\\
       \email{\{nrl39, msc252, jpp9, de226\}@cornell.edu}
}

\maketitle \begin{abstract} Medical research is one area where collecting data
is usually hard and expensive. With the launch of ResearchKit, Apple and Sage
Bionetworks made large-scale personal data collection increasingly popular via
simple text-based survey apps running on mobile phones. But such surveys can
be a barrier in terms of usability and richness of the data being collected.
In this paper, we present \sdlrx, a powerful software library designed for
ResearchKit that enables study-specific, personalized, and rich visual
surveys, for both iOS and Android platforms.

\end{abstract}





\section{Introduction}
As mayor of New York City, one of Mike Bloomberg's favorite mottos was ``In God we trust. Everyone else, bring data''.

Data today is being used for so many applications spanning across every industry.
Medical research is one area where collecting data is usually hard and expensive because of the very personal nature of the data and the need for large enough cohorts. Moreover, the increasing prevalence of chronic diseases \cite{Pollak2011-mq,Consolvo2003-xv} has increased the need for capturing data outside of clinical settings.  Because mobile phones have become truly ubiquitous , leveraging them to help collect these data has become the natural thing to do \cite{noauthor_undated-ze,noauthor_undated-hv}.  To this end Sage Bionetworks and Apple announced ResearchKit in 2015 to systematize the consent, collection, and sharing of health-relevant data.  We co-developed ResearchStack in 2016 to extend the capability to Android users. The emergence of this common software framework provides a vector for rapid iterative development, deployment, and dissemination of new techniques for data collection and interpretation.  In this paper, we present \sdlrx{}, a powerful extension to ResearchKit and ResearchStack that enables study-specific, personalized, visual surveys, for both iOS and Android platforms.

\section{Mobile Personal Data Collection Tools }
Mobile data collection is not new, in fact researchers have been using mobile phones to collect data since the early 2000s (e.g. \cite{Center_for_Disease_Control_and_Prevention_undated-ik}). In 2009, the Open Data Kit project (ODK) created an ``open-source suite of tools that helps organizations author, field, and manage mobile data collection solutions'', with a goal to ``make open-source and standards-based tools which are easy to try, easy to use, easy to modify and easy to scale.'' \cite{noauthor_undated-tj}. The tool was primarily used as a replacement for paper forms, with surveyors entering data about the local environment or asking questions to people. The tool was not primarily designed or used for people to enter their own personal information. ODK was designed for field workers, professionals, and community members to collect systematic data about people and the physical world.  Mobile Health (mHealth) tools, such as ResearchKit and ResearchStack, are intended for individuals to use in the course of their everyday lives to capture data about themselves.

With the launch of ResearchKit in March 2015, Apple made systematic personal data collection scalable: ``using smartphones to gather health data from millions of people, with their consent'' and a way to ``open a window to new insights into diseases, treatments and lifestyle effects.'' \cite{Apple_undated-rz}

``ResearchKit is an open-source software framework developed by Apple to aid clinical researchers and healthcare organizations in collecting medical information on patients and participants straight from their iPhone or Apple watch.'' \cite{Bionetworks_undated-yw}. The framework makes it easy for developers to write intuitive and standardized data-collecting mobile applications where the nature of the data collected along with the purpose of the collection are made clear to the user and translated into a corresponding scientifically validated consent form the user can understand and choose to approve, or not. The framework provides a seamless integration with HealthKit, a standardized on-device store for health and fitness data coming from the phone or from connected devices, e.g. heart rate monitor, pedometer, etc.
It also address issues such as (a) lack of standardized data, (b) lack of universal system for sharing between people, (c) app fragmentation and (d) privacy and security, as mentioned in \cite{noauthor_undated-ej}.

Restricting such data collection to iPhone users creates an obvious bias, or as stated by Deborah Estrin in the New York Times,  ``you can't just do research studies on people who can afford iPhones'' \cite{Apple_undated-rz}. In the US, 53\% of mobile phone users are Android users.
To address this issue, CornellTech and its partners developed ResearchStack, the counterpart of ResearchKit for the Android world. 

Since the launch of ResearchKit, dozens of apps have been designed deployed. These apps focus on medical research (see \cite{noauthor_undated-xa} and \cite{Tangmunarunkit2015-hu} for a list of apps), and use a small inventory of sensor-based measures, and a much larger collection of self-report surveys. These surveys are a critical aspect of all ResearchKit and ResearchStack studies, yet to date the surveys are all relatively-rigid text based surveys.   The contribution of this paper is the introduction of a visual self report technique that can be integrated into any ResearchKit and ResearchStack study to improve the usability and fidelity of data collection.

Conditions studied include: asthma, cancer (breast, skin), hepatitis C, HIV, pulmonary disease, heart disease (atrial fibrillation), multiple sclerosis, epilepsy, sports injury (concussion, torn ligaments), arthritis, microbiome,,, nutrition, autism, depression, and Alzheimer's.

\section{Visual Self-Report}
Restricting data collection to text-based multiple choice questions is a barrier both in terms of users (reading small fonts and typing on a phone can be challenging for certain populations, not to mention literacy concerns) and in terms of the data being collected (a picture is worth a thousand words). And the high quality displays and touch screens  available on today's smartphones provides a ubiquitous opportunity to move beyond text.

An alternative is visual reporting pioneered by the Cornell Information Sciences \cite{Aung2016-rx} and Small Data Lab \cite{Yang2016-dn}, \cite{noauthor_2015-gf} , where personalized images are used to improve survey efficiency.
As pointed in \cite{Yang2016-dn}, ``asking generic sets of questions repeatedly introduces user burden and fatigue that threatens to interfere with their utility''. And using images ``offers several potential benefits: both broader and more specific coverage of activities of daily living, improved engagement, and accurate capture of individual health situations''.

\begin{figure}[h!]
\centering
\includegraphics[width=8cm]{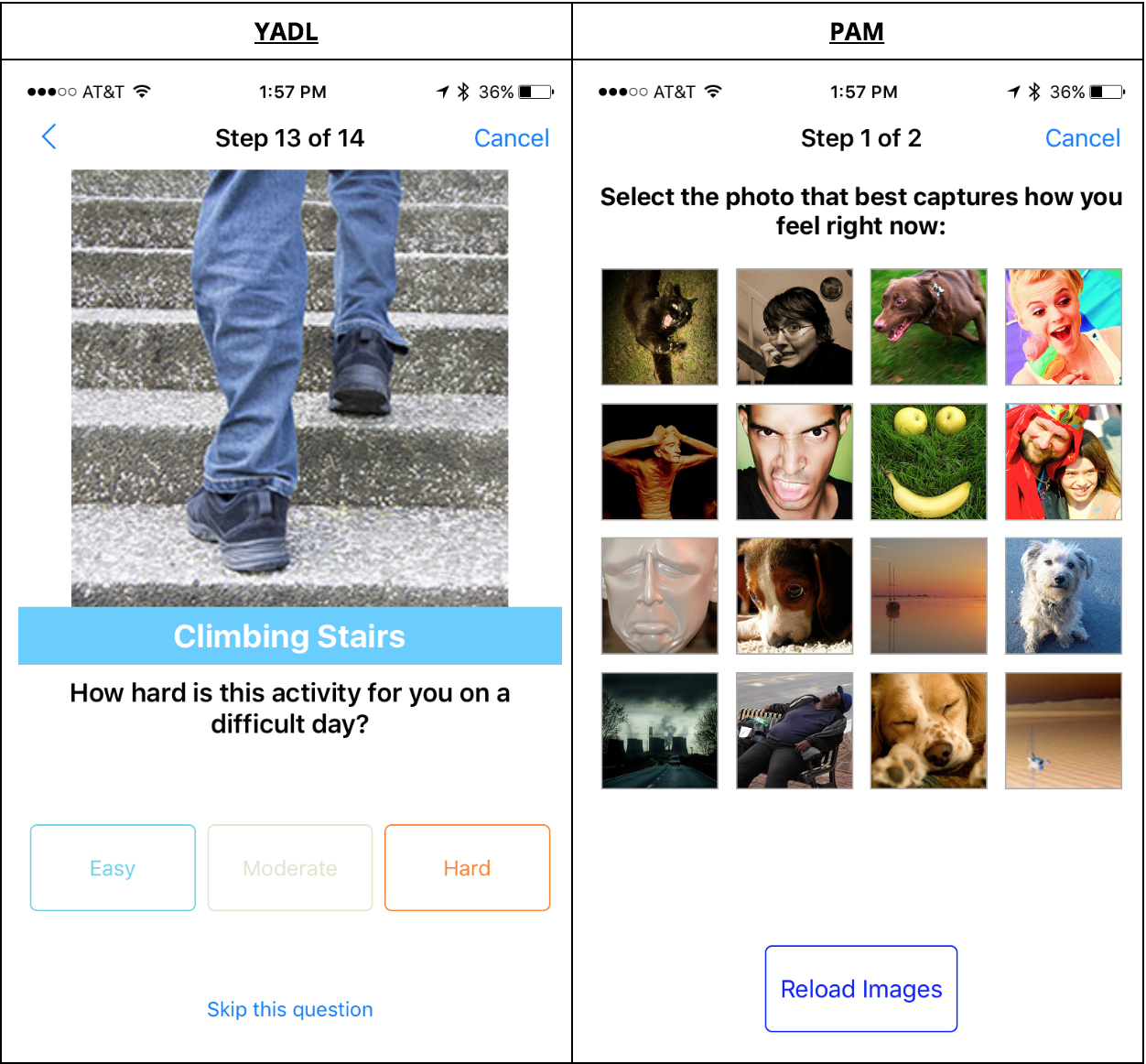}
\caption{two examples of visual reporting apps.}
\label{fig::visual-apps}
\end{figure}

Another interesting aspect of visual surveys is that they can be implemented as a two step process where users start by selecting from a large pool of items (e.g. once a month) and only need to interact with the selected ones on a daily basis. This can apply to penible activities, medications taken, etc. Figure~\ref{fig::visual-apps} features YADL (for daily activity reporting) and PAM (for mood reporting).

Both PAM and YADL were implemented as standalone applications for both iOS and Android platform, without much code reuse. They both required some programming expertise and a non-negligible software engineering budget.

\section{\sdlrx}
\sdlrx{} is a powerful software library designed for ResearchKit that enables study-specific, personalized, and rich visual surveys, for both iOS and Android platforms. Researchers can easily incorporate visual self-report mechanisms into mobile apps built with ResearchKit and ResearchStack. We want to emphasize the fact that we are trying to lower the barrier of entry for building such apps, both in terms of expertise and cost.

Rather than reinventing the wheel and creating more fragmentation (see \cite{noauthor_undated-ej}), the goal for \sdlrx{} was to build on top of ResearchKit/Stack, to enrich existing application with more intuitive way of collecting user data. Therefore, our guiding principles leant towards an architecture that is (a) modular, (b) opinionated and (c) that favors compatibility over creativity.

\begin{figure}[h!]
\centering
\includegraphics[width=8cm]{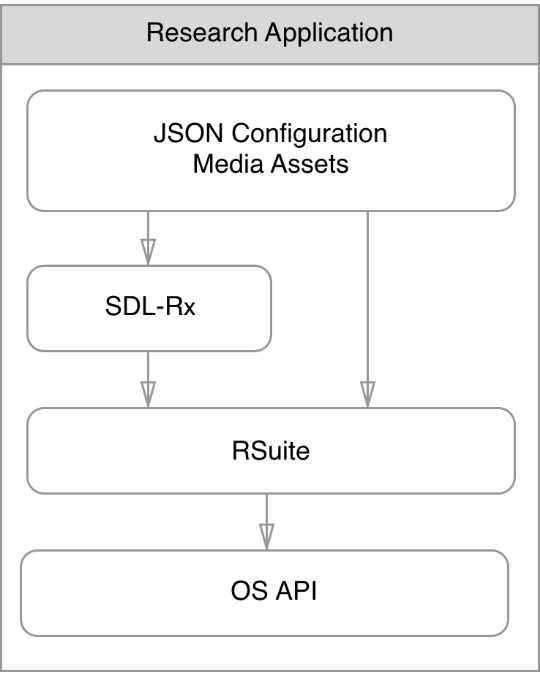}
\caption{architecture diagram.}
\label{fig::architecture}
\end{figure}

Using \sdlrx{}, an application is defined by a JSON file that defines the structure of the app and references some media assets to be used. The JSON file for the YADL example from Figure~\ref{fig::visual-apps} is presented in Figure~\ref{fig::json}.

\lstset{
basicstyle=\small\ttfamily,
columns=flexible,
breaklines=true
postbreak=\raisebox{0ex}[0ex][0ex]{\ensuremath{\color{red}\hookrightarrow\space}}
}
\begin{figure*}[h!]
\begin{Verbatim}[fontsize=\small,frame=single]
{
   "YADL":{
      "full":{
         "identifier":"YADL Full Identifier",
         "prompt":"How hard is this activity for you on a difficult day?",
         "summary":{
            "identifier":"YADL Full Summary Identifier",
            "title":"Thanks",
            "text":"Thank you for participating in the YADL Full Assessment"
         },
         "choices":[
            {
               "text":"Easy",
               "value":"easy",
               "color":"#69D2E7"
            },
            {
               "text":"Moderate",
               "value":"moderate",
               "color":"#E0E4CC"
            },
            {
               "text":"Hard",
               "value":"hard",
               "color":"#F38630"
            }
         ]
      },
      "spot":{
         "identifier":"YADL Spot Identifier",
         "prompt":"Which activities did you have trouble with today?",
         "summary":{
            "identifier":"YADL Spot Summary Identifier",
            "title":"Thanks",
            "text":"Thank you for participating in the YADL Spot Assessment"
         },
         "noItemsSummary":{
            "identifier":"YADL Spot No Activities Summary Identifier",
            "title":"Thanks",
            "text":"You have no activities to measure"
         },
         "options":{
            "somethingSelectedButtonColor":"#0080ff",
            "nothingSelectedButtonColor":"#FCC103",
            "itemCellSelectedColor":"#7FEE7D",
            "itemCellSelectedOverlayImageTitle":"first_tab",
            "itemCollectionViewBackgroundColor":"#E9E9E9",
            "itemsPerRow":3,
            "itemMinSpacing":10.0
         }
      },
      "activities":[
         {
            "imageTitle":"Bathing",
            "description":"Bathing",
            "identifier":"Bathing"
         },
         {
            "imageTitle":"BedToChair",
            "description":"Bed To Chair",
            "identifier":"BedToChair"
         },
[...]
         {
            "imageTitle":"Toilet",
            "description":"Using the toilet",
            "identifier":"Toilet"
         },
         {
            "imageTitle":"WalkingUpStairs",
            "description":"Climbing Stairs",
            "identifier":"WalkingUpStairs"
         }
      ]
   }
}
\end{Verbatim}
\caption{JSON configuration file for the YADL application.}
\label{fig::json}
\end{figure*}

\sdlrx{} is totally storage-agnostic. Sending the data to the cloud and storing it is out of the scope of the framework. Existing solutions such as Sage Bionetworks Bridge server \cite{noauthor_undated-ax} or Small Data Lab Ohmage server \cite{Software_undated-mu} can be used easily as back-ends.

\section{Examples of app (soon to be built)}
We provide three use cases currently in development that demonstrate the broad applicability and customizability of the approach:

Dr. Fred Muench of Northwell Health, the largest integrated health system in New York, is developing a research study on impulsivity that incorporates several validated measures as active tasks, collects passive data streams to inform the definition of novel digital biomarkers, and will use Visual self report to more easily document their personal impulsivity associated behaviors that are not measurable through sensors: eating, alcohol consumption, shopping, etc..

Dr. No\'{e}mie Elhadad at Columbia University is conducting a set of research studies about endometriosis under an umbrella project called `Citizen Endo'. Symptom and trigger monitoring is critical to understand this little understood and under-diagnosed reproductive disease and improve treatment. The pain associated with endometriosis interferes with many activities of daily living and the group are exploring the use of YADL to improve the usability and fidelity of participant self report around symptoms and interfered activities, including, work, daily routines, sex, and exercise.

Dr. Vijay Vad of Hospital for Special Surgery  is using the Visual self Report technique for Lower Back Pain patients to self report on recovery progress in response to home exercise programming and we plan to incorporate this into a ResearchKit and ResearchStack study in the future.

\section{Future Work}
Our first priority is to have researchers use our framework to build and deploy personal data collection applications. With \sdlrx{}, we tried to lower the bar (both in terms of cost and expertise). But there is still a lot to do. The end goal is to build such apps as easily as creating a survey on platforms such as Google Forms, Qualtrics or Survey Monkey.

We want to provide ready-to-use ResearchKit and ResearchStack-compatible components to capture more data. This implies leveraging and integrating with location, phone movement, phone interaction, image capture, etc. Some interesting apps have already been built but not out of reusable components.

We want to improve the orchestration of such surveys. It is important to ask the right question the right way. It is even more important to ask the question at the right time. By orchestration, we include schedule (when), frequency (how often) and reminders (e.g. ability to snooze a survey). All of these features should be simple parameters a researcher can tweak for a given research study.

We are actively involved in the development of new digital biomarkers based on active and passive  sensor data collection to reduce dependence on self report. Progress is critically dependent on improved self-report such as \sdlrx{} to provide labeled data sets.

We also want to explore and encourage similar data collection apps beyond purely medical research studies. Quality-of-life issues including pollution (noise, air) and happiness in cities would be a good first start.

\section{Conclusion}
Mobile phones have become one of our most personal properties. It makes sense to leverage them to let users capture their personal data for research studies. Apple and Sage Bionetwork's ResearchKit made a big contribution by creating a platform to build such personal data collection app, using traditional survey forms. With \sdlrx{}, we enrich the ecosystem with ready-to-use components to incorporate visual surveys into ResearchKit and ResearchStack for iOS and Android platforms, respectively.

We have been using \sdlrx{} internally at Cornell Tech for various projects. We encourage the community to use it and provide comments. The Small Data Lab ResearchKit Extensions (sdl-rkx) and Small Data Lab ResearchStack Extensions (sdl-rsx) packages, available immediately as open source projects under the Apache 2.0 license on the Cornell Tech public github repository.

A demo app that leverages the extension is included in each repository:
\begin{itemize}[noitemsep, topsep=-5pt]
\item \href{https://github.com/cornelltech/sdl-rsx}{https://github.com/cornelltech/sdl-rsx}
\item \href{https://github.com/cornelltech/sdl-rkx}{https://github.com/cornelltech/sdl-rkx}
\end{itemize}

\bibliographystyle{ieeetr}
\vspace{2.5mm}
\bibliography{references}

\end{document}